\documentclass[english]{article}
\usepackage[LGR,T1]{fontenc}
\usepackage[latin9]{inputenc}
\usepackage{graphicx}
\usepackage{pdfpages}

\makeatletter
\newcommand{\lyxaddress}[1]{
	\par {\raggedright #1
	\vspace{1.4em}
	\noindent\par}
}

\usepackage{babel}
\date{}

\makeatother

\usepackage{babel}
\begin{document}
\title{Quantifying the performance of high-throughput directed evolution
protocols}
\author{Adèle Dramé-Maigné{*}, \and Anton Zadorin{*},\and Iaroslava Golovkova,\and
Yannick Rondelez}
\maketitle

\subsubsection*{}

\lyxaddress{Laboratoire Gulliver, CNRS, ESPCI Paris, PSL Research University,
10 rue Vauquelin, Paris, France. {*} equally contributed to this work.}

\subsection*{Abstract}

Most protocols for the high-throughput directed evolution of enzymes
rely on random encapsulation to link phenotype and genotype. In order
to optimize these approaches, or compare one to another, one needs
a measure of their performance at extracting the best variants. We
introduce here a new metric named the Selection Quality Index (SQI),
which can be computed from a simple mock experiment with a known initial
fraction of active variants. As opposed to previous approaches, our
index integrates the random co-encapsulation of entities in compartments
and comes with a straightforward experimental interpretation. We further
show how this new metric can be used to extract general trends of
protocol efficiency, or reveal hidden mechanisms such as a counterintuitive
form of beneficial poisoning in the Compartmentalized Self-Replication
protocol.

\subsection*{Introduction}

Molecular Directed Evolution (DE) is a technique to obtain biomolecules
with desirable or improved functions, by iteratively generating pools
of randomized variants and screening or selecting them to extract
the best performers from these pools. This technique, initially developed
for the search of nucleic acid ligands\cite{tuerk_systematic_1990,ellington_vitro_1990},
has been extended to proteins and used to explore other chemistries
and functions \cite{chen_tuning_1993,arnold_directed_2018}. Most
importantly, it is now also used to select biopolymers with tailored
catalytic properties, i.e. artificial enzymes\cite{arnold_directed_2003}.

The first high-throughput enzymes DE experiments used living cells
as a selection medium for the mutant phenotypes. In such \emph{complementation}
approaches, the cell expression machinery is employed to translate
an exogenous randomized genetic element into the corresponding mutated
polypeptide. Additionally, survival and replication of the host cells
is made dependent on the catalytic activity provided by the exogenous
gene. Selection between desirable and undesirable phenotypes becomes
possible for two reasons: first, the cell boundaries ensure that the
genotype-phenotype linkage is maintained, and that selection, acting
on phenotypes, yields improved transferable genotypes; second, cell
transformation strategies provide a straightforward way to obtain
populations of cells where each member contain one and only one of
the possible mutant genotypes. Alternatively, when no strong coupling
between the exogenous phenotype and host cell survival can be achieved,
screening methods, where the cells can be sorted according to an observable
marker of phenotype, have been designed.

While these \emph{in vivo }implementations are very efficient, they
still suffer from a number of technical issues. First, complementation
approaches are only possible for catalytic activities that are related
to the host cell's essential processes. In the screening protocols,
permeability of the cell walls to metabolite may interfere with the
diffusion of substrate to the catalyst. Finally, in both cases, the
strategy has to be compatible with the metabolic function of the living
host; activities that are toxic for the host gets naturally counter-selected,
in some case disrupting the intended DE process.

To bypass these problems, a number of recent developments have introduced
the idea of performing high-throughput directed enzyme evolution \emph{in
vitro}\cite{ghadessy_directed_2001,ellefson_directed_2014,griffiths_directed_2003,baret_fluorescence-activated_2009,kintses_picoliter_2012,colin_enzyme_2015,yamauchi_evolvability_2002,povilaitis_vitro_2016,agresti_ultrahigh-throughput_2010,beneyton_cota_2014,fallah-araghi_completely_2012}.
In these protocols, mutant genotype-phenotypes are randomly distributed
in artificial micro-compartments, with an internal volume ranging
between the femtoliter and the nanoliter. As is the case for DE using
living organisms carriers, two strategies can be used to extract the
best genotypes: screening and selection. In screenings, the phenotypes
induce a detectable modification of the physico-chemistry of the compartment,
which is used to sort the compartments one-by-one into desirable and
non-desirable bins\cite{baret_fluorescence-activated_2009}. In selection
approaches, the artificial compartment additionally enables a biochemical
genetic amplification process, directed at the gene of interest and
whose activation is conditional on the desired activity.

A common important feature of these high-throughput \emph{in vitro}
protocols is that they use a \emph{random} distribution of the mutant
library in the artificial micro-compartments (usually droplets, but
also liposomes, micro-patterned arrays\cite{rondelez_microfabricated_2005},
etc). By contrast with the case of \emph{in vivo} approaches, \emph{in
vitro} protocols provide no internal mechanism to insure that one
and only one genotype will be present in each compartment. As a consequence,
one can only control the \emph{average} number of mutants per compartment
(a value referred as $\lambda$ in the following discussion) but not
their actual repartition in the compartment population. If the partitioning
process is truly random, the distribution of occupancy will follow
the Poisson law. For example, if $\lambda$ is set to 1 (i.e. there
are as many compartments as mutants genotypes), the proportion of
compartment containing 0, 1 and more than one genotypes will be roughly
one third each. Lowering the values of $\lambda$ yields more wasted
(unoccupied) compartments, while higher values increases the number
of compartments containing multiple genotypes. 

High-throughput protocols are therefore fundamentally stochastic,
and their purpose is to increase the proportion of highly functional
variants after each round. Their performance at enriching libraries
will depend on a number of experimental choices or constraints. These
factors include the value of $\lambda$, but also, for example, the
contamination by parental genotype, the homogeneity of the compartment's
physicochemical parameters, or the robustness of the genotype-phenotype
linkage, etc. Various experimental approaches have been proposed,
with different throughput and performance. Assessing the relative
efficiency of the methods and the particular contribution of these
factors on a given DE protocol can be challenging. 

\subsubsection*{Enrichment factor}

\begin{figure}
\begin{centering}
\includegraphics[width=1\textwidth]{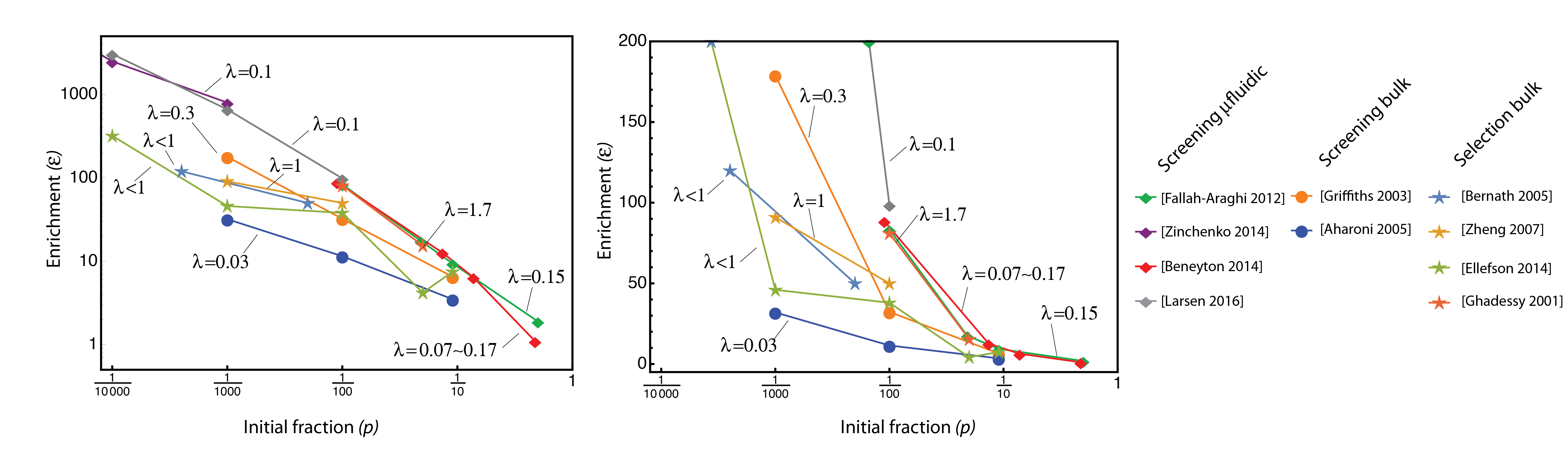}
\par\end{centering}
\caption{\label{fig:enrich biblio}Reported enrichment factors
from a selection of references. The enrichment factor is plotted in
linear (right) or log (left) scale against the initial fraction of
active mutants used in the experiment. Lines connect data points from
the same manuscript, and obtained with a single value of $\lambda$
(indicated next to the line). Disks are used for protocol using screening
and polydisperse emulsions; diamonds, for protocols using screening
and monodisperse (microfluidic) emulsions; Stars, for protocols using
selections in polydisperse emulsion. Some values were recomputed from the reported date. See SI for details.}
\end{figure}

In an attempt to characterize the quality of a given high-throughput
directed evolution protocols, many papers report a so-called enrichment
factor $\varepsilon=\frac{p'}{1-p'}\cdot\frac{1-p}{p}$, where $p$
and $p'$ are the frequencies of the allele of interest before and
after one selection cycle, respectively (in some cases the enrichment
factor is defined using the simple ratio of frequencies, $\varepsilon=p'/p$).
To evaluate the value of this enrichment factor for a given protocol,
an experiment is set up where the initial library contains a small
proportion of active variants among many inactive ones, and this proportion
is measured again after a single selection round. Generally, to perform
these experiments, a mock library containing only two versions of
the target protein, an active and an inactive one, is used. A compilation
of reported values of $\varepsilon$ from various reports is given
in Fig \ref{fig:enrich biblio}, while the full tables are included
in SI Material \& Methods (Table S2, S3, S4).

However, the usage of $\varepsilon$ as a measure of quality comes
with a number of issues: 
\begin{itemize}
\item First, $\varepsilon$ depends strongly both on the initial fraction
of functional mutants $p$ and on the value of $\lambda$. As can
be seen in Fig \ref{fig:enrich biblio}, measurements of enrichment
factors made with different values of $p$ generally yield values
of $\varepsilon$ that vary over many orders of magnitude. Similarly,
changes in $\lambda$ yield different enrichment factors, with no
clear interpretation. 
\item Second, the upper theoretical limit of $\varepsilon$ is infinity
when $p$ or $\lambda$ become small. This reflects the fact that
in principle, a flawless selection or screening assay should get rid
of the poor phenotype in a single round. It is therefore difficult
to estimate what should be a correct, or acceptable, value of an experimental
$\varepsilon$. In practice, smaller values of $p$ or $\lambda$
tend to increase the observed value of $\varepsilon$, suggesting,
against common knowledge, that the protocol performs better on less
rich libraries. Additionally, measurements at very small $\lambda$
are not really useful in assessing a protocol meant to provide the
highest possible throughput, as the throughput is actually proportional
to $\lambda$. Note that the theoretical lower limit of $\varepsilon$
is 0 at high $p$ or high $\lambda$. Every value below 1 would indicate
a process that counter-selects for the good genotype. 
\end{itemize}
Altogether, interpreting the enrichment factor is difficult and it
is clear that one cannot use it directly to compare different protocols
in terms of their ability to enrich libraries. We introduce here an
improved performance metrics, the Selection Quality Index (SQI). The
SQI is expressed as a fraction of the theoretical (achievable) value
for the selection of a functional variant from a pool of lethal variants
and normalizes for the effect of $p$ and $\lambda$. An SQI of 1
denotes a perfect protocol, whereas an SQI of 0 indicates a complete
failure, i.e. the absence of any enrichment effect (negative values
indicate counter-selection). Accordingly, this index -in combination
with the throughput of the protocol- can be used to assess the true
absolute potential of a given experimental approach to concentrate
rare catalytic variants. It can also be used to compare different
protocols, or different variants of the same protocol.

\subsubsection*{Model}

The factors that control the change in frequency of a mixture of active
and inactive mutants in a high-throughput protocol can be separated
in two groups.

The first group contains controllable or known parameters of the experimental
design: $\lambda$, the sharing status, and the replication/selection
function. These effects are detailed below: 
\begin{itemize}
\item As mentioned above, when $\lambda$ increases, a fair fraction of
the compartments actually contain more than one phenotype, leading
to the possibility of ``hitch-hiking''. Inactive mutants, which
should not have survived the screening or selection process, are carried
over by sharing a compartment with a better phenotype. In principle,
knowledge of the compartmentalization process allows us to describe
the distribution of occupancies. For example, in the case of random
compartmentalization, this distribution will follow a Poisson law. 
\item The second effect, also linked to the random co-encapsulation of variants,
is fitness sharing. This occurs when the presence of the poor phenotype
is actually detrimental to the good one. This is typically the case
in selection experiments: an equal fraction of the replication potential
associated with one compartment will be used to amplify each local
variant, whether or not they actually contributed to the replication
activity. This sharing effect does not happen in screening protocols,
which works by discarding the low activity compartments, rather than
replicating the genotypes in the good ones. In other words, in screenings,
a variant present in a supra-threshold compartment contributes one
genotype to the next generation irrespective of the presence of co-encapsulated
variants in his compartment. These two situations are refereed as
sharing and non-sharing, respectively, in the rest of the manuscript. 
\item Third, the replication/selection function\cite{zadorin_selection_2017},
which describes how many genotypic copies are pushed to the next generation,
according to the phenotypic activity observed in a compartment. In
screenings for example, the value of the function is one if the phenotypic
activity is above a given threshold, and 0 otherwise. In selections,
it is set by the internal replication chemistry.
\end{itemize}
The second group consists of experimental idiosyncrasies that are
not directly controllable. Together, they explain the observed deviation
from a perfect assay. These factors can be based on a variety of causes,
some known or guessable, other altogether unknown. For example, in
screening test, the sorting machinery may not be 100\% efficient or
generally makes a number of errors (typically increasing when sorting
is attempted at higher frequencies \cite{baret_fluorescence-activated_2009}).
Noise in expression can also play a role when compartments containing
the same genotype may end up with various level of phenotypic activity
(typical when bacterial expression systems is used, i.e. lysate assays
\cite{kintses_picoliter_2012}). Additionally, leaks between compartments,
poor genotype-to-phenotype linkage (e.g. in mRNA display strategies),
hidden selection biases (e.g. linked to toxicity effect), non-homogeneity
or polydispersity of the encapsulating compartments (see below), carry-over
of the parent genomes in selections, and a myriad of other causes
may affect the result. One typically does not have enough information
to precisely describe or model these effects, and is rather interested
in quantifying their collective negative impact on the selection protocol,
in order to empirically try to minimize it.

\subsubsection*{The Selection Quality Index (SQI)}

We have recently shown that all effects belonging to the first group
can be modeled using the single equation (\ref{eq: zadorin2017}).
This equation describes the evolution of a population of variants
submitted to random-compartmentalized selection rounds, characterized
by an additive replication/selection function $f$, a sharing rule
$\varphi(n)$ (the fraction of the value given by $f$ allocated as
copies to each individual in the droplet of size $n$), and a distribution
of phenotypes $\rho$ (\cite{zadorin_natural_2017}).

\begin{equation}
\rho'=\frac{{\displaystyle \sum_{n=0}^{\infty}\frac{\lambda^{n}}{n!}\varphi(n+1)\langle\delta_{x}*\rho^{*n},f\rangle}}{{\displaystyle \sum_{n=0}^{\infty}\frac{\lambda^{n}}{n!}\varphi(n+1)\langle\rho^{*n+1},f\rangle}}\,\rho\label{eq: zadorin2017}
\end{equation}

\noindent where$*$ means convolution of distributions, $\delta_{x}$
is the delta-function centered at $x$, and $\langle\rho,f\rangle$
means application of the distribution $\rho$ to the function $f$.

Here, we are interested in quantifying the quality of a typically
mock selection protocol where only two variants are used. One of them
is active and the second completely dysfunctional (i.e., in isolation,
it should not survive the selection process). Therefore $\rho$ has
only two discrete values.
\begin{figure}
\begin{centering}
\includegraphics[scale=0.6]{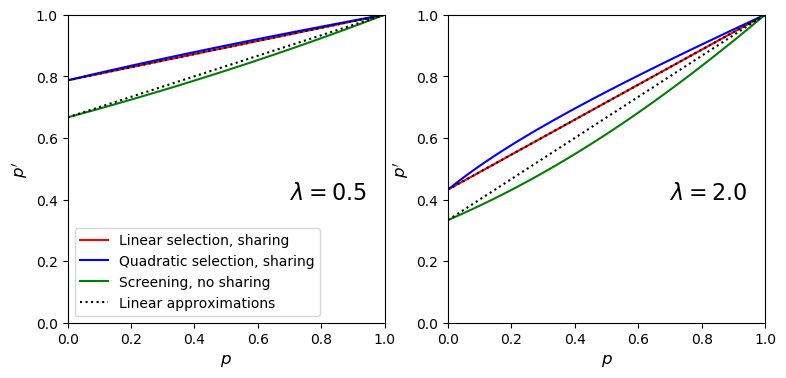}
\par\end{centering}
\caption{\label{fig:p'=00003Df(p)} Graphical representation of the expected
new fraction $p'$ as a function of initial fraction $p$ for various
replication/selection function $f$ and for two values of $\lambda$.
The exact analytical results derived from (\ref{eq: zadorin2017})
are shown by solid lines. The linear approximation is shown by a dotted
line for the cases both with and without sharing. $\Delta$ corresponds
to the intercept of the curves with the vertical axis and is independent
of \emph{f}. }

\end{figure}
Equation \ref{eq: zadorin2017} allows us to compute the theoretical
(expected) value of $p'$ as a function of $p$ for different $\lambda$,
which we will call $p'_{theo}(p)$ (Fig. \ref{fig:p'=00003Df(p)}).
Importantly, this function is discontinuous at $p=0$, reflecting
the effortless invasion of the population of dysfunctional mutants
by a rare functional one. In an ideal selection with no co-encapsulation,
the jump observed at low $p$, corresponding to the selection of one
active variant among an infinity of inactive ones would be equal to
1. Therefore, the value of this jump, $\Delta$, provides a good absolute
characterization of the selection efficiency at low $p$, and depends
only on $\lambda$. Indeed, one can show that the general equation(\ref{eq: zadorin2017})
yields an expression for $\Delta$, which does not depend on the replication
function $f$:

\[
\Delta=p'_{theo}(p\rightarrow0)=\frac{{\displaystyle \sum_{n=0}^{\infty}\frac{\lambda^{n}}{n!}\varphi(n+1)}}{{\displaystyle \sum_{n=0}^{\infty}\frac{\lambda^{n}}{n!}(n+1)\varphi(n+1)}}.
\]

However, it is generally not practical to perform experiments at very
small $p$, so we will also need the behavior of $p'$ at $p\rightarrow0$.
As this behavior does depend on $f$, we propose to correct using
a linear approximation of the following form. 

\begin{equation}
p'_{theo}(p)\simeq\Delta+(1-\Delta)p.\label{linear-approx}
\end{equation}

This expression is exact for linear replication functions and performs
well for nonlinear ones (especially, for not very large $\lambda$,
see Fig. \ref{fig:p'=00003Df(p)}). Altogether, we find that all possible
scenario collapses into only two cases:

In the non-sharing situation, which is typical of screening protocols,
we have $\varphi(n)=1$ and the theoretical value of $\Delta$ as
a function of $\lambda$ is given by:

\begin{equation}
\Delta=\frac{1}{1+\lambda}.\label{screen-non}
\end{equation}

In the sharing situation, which is more typical of selections assay,
there is a cost associated with the presence of co-encapsulated variants
and $\varphi(n)=1/n$. We obtain:

\begin{equation}
\Delta=\frac{1-e^{-\lambda}}{\lambda}.\label{lin-shar}
\end{equation}

We conclude that, knowing $\lambda$ and $p$, and the sharing behavior
of a given experiment setup, one can obtain a good approximation of
the theoretically expected gene frequency $p'_{theo}(p)$ after one
round of selection of a very rare functional mutant in a population
of dysfunctional ones. This $p'_{theo}$ can be used as a reference
to assess the quality of the experimental protocol, independently
of the controllable parameters $\lambda$ and $p$. In principle,
one can thus use any convenient value of $\lambda$ and $p$ and simply
measure the experimental frequency $p'_{exp}$ after the selection
cycle. The SQI of the protocol can then be expressed as the ratio
of the experimental to theoretical frequency jumps, i.e:

\begin{equation}
SQI=(p'_{exp}-p)/(p'_{theo}-p).\label{eq:sqi}
\end{equation}

Finally,

\[
SQI_{screen}=\frac{(1+\lambda)(p'-p)}{1-p}\quad and\quad SQI{}_{selection}=\frac{\lambda e^{\lambda}(p'-p)}{(e^{\lambda}-1)(1-p)}.
\]

\noindent These formula provide an absolute measurement of protocol
quality and selection efficiency, that corrects for $p$, $\lambda$
and the sharing behavior. It is independent of the replication/selection
function $f$, and reflects only the experimental contingencies belonging
to group 2.

\subsubsection*{General trends for SQI values}

\begin{figure}
\begin{centering}
\includegraphics[width=1\textwidth]{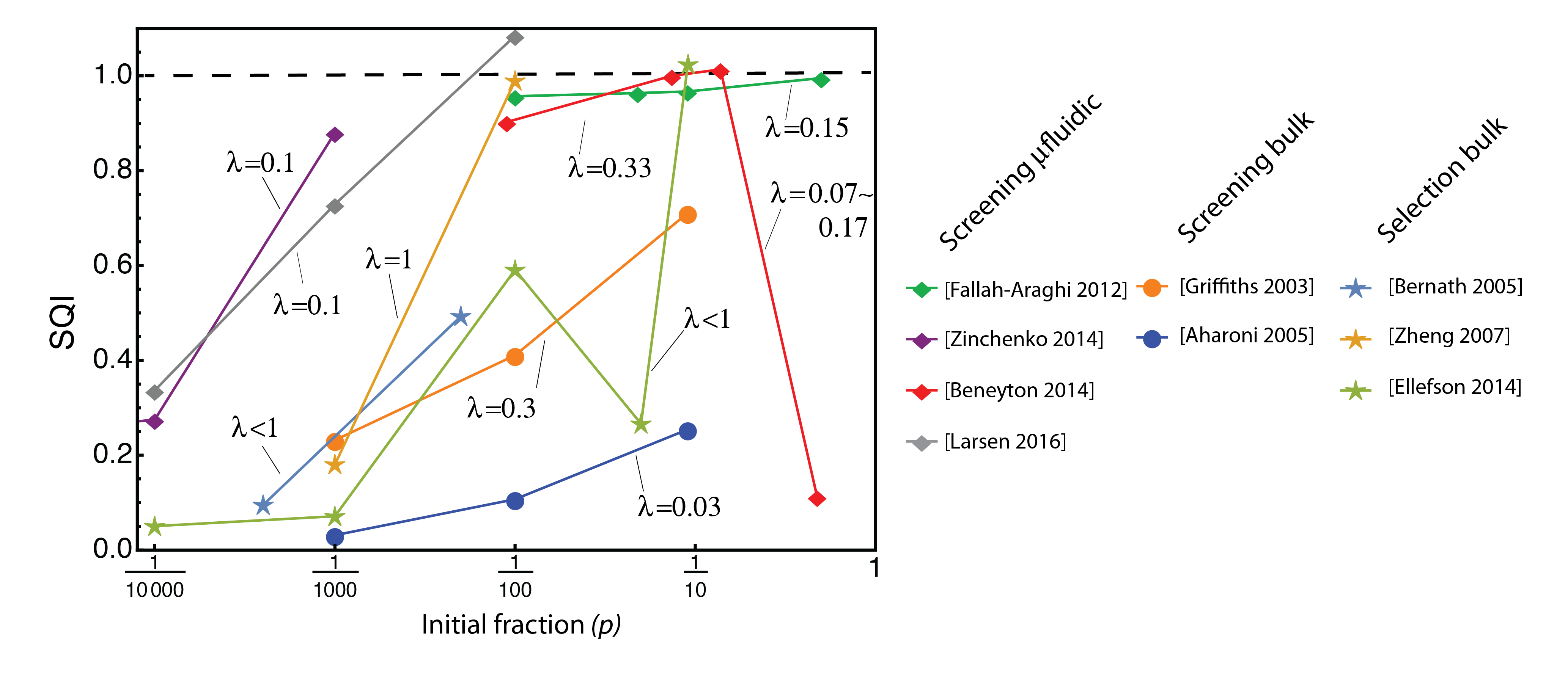}
\par\end{centering}
\caption{\label{fig:Pi-scores}SQI calculated for the data in Fig \ref{fig:enrich biblio}.
Lines connect data point originating from the same manuscript, and
done with a single value of $\lambda$ (indicated next to the line).
Disks are used for protocol using screening and bulk (non-monodisperse)
emulsions; diamonds, for protocols using screening and microfluidic
(monodisperse) emulsion; Stars, for protocols using selections in
bulk emulsion. The dotted line at $=1$ represents the theoretical
maximum performance, once initial fraction and random partitioning
have been taken into account. Note that the CSR experiment\cite{ghadessy_directed_2001} is off scale here (see Fig\ref{fig:Pi-scores-exp}).}
\end{figure}

In Fig. \ref{fig:Pi-scores} we use the reported data shown in Fig.
\ref{fig:enrich biblio} to compute the SQI of a variety of experimental
protocols and conditions. This immediately reveals some interesting
trends. 

First, some of the data series, taken at various but relatively high
initial fractions (e.g. for $p\geq1/100$) collapse to a single SQI
value. For example the points from {[}Fallah-Araghi 2012{]} and {[}Beneyton
2014{]}, now give roughly the same SQI, independent of the initial
fraction. Additionally, for these relatively high initial fractions,
we note a cluster of data points that achieve an SQI close to one,
indicating that the best achievable enrichment performance has been
attained. 

Second, for the lower $p$ values, the selection efficiency always
decrease with decreasing initial fractions of active mutants. This
stands in striking contrast with the trend observed in Fig. \ref{fig:enrich biblio},
where enrichment factors seemed to indicate that all assays performed
much better at lower initial fraction. This apparent trend was just
a mechanical consequence of the fact that it is easier to invade a
population of dysfunctional mutants, rather than a population already
containing a significant fraction of functional mutants. Our analysis
corrects for that effect. It thus highlights the fact that other experimental
contingencies actually dominate at low $p$, making experimental protocols
less and less ideal as the initial fraction decreases. This is due
to a decreasing signal-to-background ratio, where the background noise
can result from a variety of effects, such as contamination by parent
DNA in selections, sorting errors in screenings, or the difficulty
to recover very small amounts of genetic material in the general case.

Third, protocols can be separated according to their type: only screening
using microfluidic compartmentalization appear to yield efficiencies
close to 1. In contrast, protocols using screening of bulk emulsions
typically have a lower performance. This indicates that the size-dispersity
of the compartments is an important factor contributing to the efficiency
of screenings. In contrast, some selection processes are able to obtain
high index without resorting to monodisperse emulsion, which could
be related to the superior robustness of selections versus screenings,
with respect to co-encapsulation (already noted in \cite{zadorin_selection_2017}).
Indeed, reports concerning selections tend to use higher $\lambda$
values. 

Besides these general cross-studies trends, one can also use the SQI
to analyze individual experimental setups. For example, In 2009, Baret
and colleagues tested the Fluorescent Activated Droplet Sorting method
\cite{baret_fluorescence-activated_2009} with different conditions,
and concluded that poisoning co-encapsulation explained the lower
efficiency at higher $\lambda$. Re-analysis of these results shows
that the SQI, although correcting for $\lambda$, decreases clearly
as well. This suggests that another cause linked to the increased
concentration of cells, e.g. leaks, could explain the behavior. 

\subsubsection*{The SQI reveals deviation to Poisson statistics}

\begin{figure}
\begin{centering}
\includegraphics[width=1\textwidth]{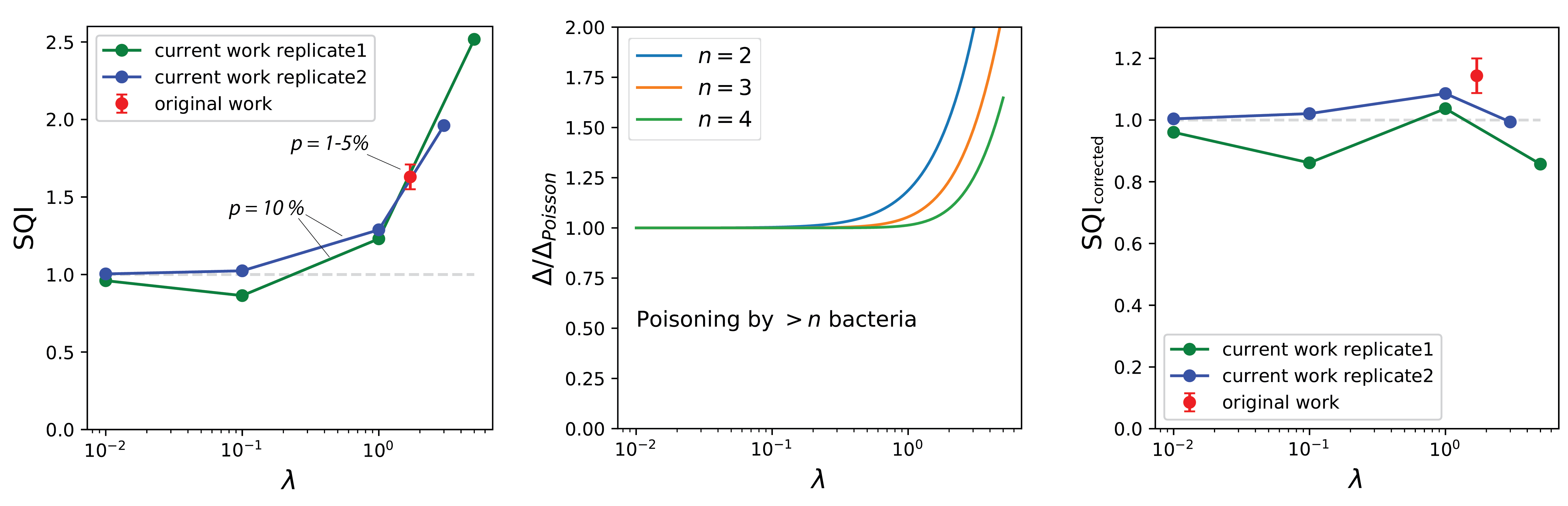}
\par\end{centering}
\caption{\label{fig:Pi-scores-exp}Left: SQI calculated for CSR protocols at
various $\lambda$. Red: original CSR report\cite{ghadessy_directed_2001}, in polydisperse emulsion.
Blue and green: this work, two independent experimental replicates
in monodisperse emulsions. The dotted line at SQI $=1$ represents
the theoretical maximum efficiency. Middle: Change in expected frequency
jump assuming that droplets containing more than n bacteria are poisoned
and do not participate in the reaction. Right: The SQI is recomputed
with a cutoff set to n = 2, corresponding to the experimental observation
that PCR is quickly poisoned by excess lysate.}
\end{figure}

The SQI seems to be able to stratify experimental platforms in terms
of both the quality of the experimental design, and their reaction
to increasingly challenging libraries (that is, containing a smaller
and smaller fraction of active variants). One approach, however, seems
to stand out in terms of its SQI, which is much above 1 when measured
for a high value of $\lambda=1.7$. This approach is an \emph{in vitro}
selection process, termed CSR, for Compartmentalized Self-Replication\cite{ghadessy_directed_2001}.
It targets a bacterial library of variants of the Taq polymerase gene,
which is encapsulated in droplet with Taq-specific primers. After
lysis at high temperature and PCR thermal cycling, mutants polymerases
with higher activity replicate their own genes better than less active
ones. When the droplets are broken, the retrieved genetic population
is thus enriched in active Taq variants. According to its SQI, CSR
apparently selects active mutants better than the theoretical limit
due to random coencapsulation (Fig. \ref{fig:Pi-scores-exp}). This
could indicate that this particular selection process (and possibly
some other) has a build in mechanism that limits the carryover of
inactive variants. To investigate this point, and since, the SQI in
this case was evaluated from relatively sparse reported information,
we decided to re-implement the experiment.

After a number of failed attempts, we were able to reproduce the CSR
with some modifications. First, we used Klentaq\cite{klenow_selective_1970},
a lysate-robust polymerase variant, instead of Taq. We found that
his enzyme was expressed at sufficient levels in KRX cells (while
the initial report used TG1 cells). And finally we used monodisperse
droplets in fluorinated oils, instead of the polydisperse emulsion
in mineral oil initially reported (see SI experimental section). Given
these adjustments, we were indeed able to observe the self-replication
of the polymerase gene, both in bulk solution, and in 24 $\mu$m droplets.
We thus measured SQI at different values of $\lambda$, as shown in
Fig \ref{fig:Pi-scores-exp}. These measurement showed that the CSR
process is indeed able to perform optimally at low $\lambda$ values,
were the SQI is roughly 1. It also confirmed that higher $\lambda$
result in SQI clearly above 1, and therefore that the CSR reaction
is somehow immune to random co-encapsulation. 

In the course of the experiments, we also noted that the self-PCR
reaction performed less efficiently when higher concentration of bacteria
were used in the master mix. This is most likely associated with the
strong toxicity of the bacterial lysate toward the PCR reaction\cite{kermekchiev_mutants_2009}.
We thus reasoned that this phenomenon could explain the observed behavior:
if droplets containing more than one bacteria are unable to support
PCR, whatever the mixture of active and inactive variants they contain,
then they simply do not contribute to the new generation. Therefore,
it is as if co-encapsulation would not occur and the jump in frequency
should be less affected (Fig. \ref{fig:Pi-scores-exp} Middle). We
therefore performed the reaction, in test tubes, with a mixture of
variants that reproduce the content of droplet containing one active,
plus increasing amount of inactive variants (see SI experimental section).
Indeed, we find that the self-PCR reaction yield decreases a lot as
soon as a single inactive variant is encapsulated with an active wild-type.
The reaction does not happen with two or more encapsulated inactive
bacteria. Fig. \ref{fig:Pi-scores-exp} right shows the recomputed
SQI taking this observation into account. In the case, the corrected
SQI stays close to 1 for all experiments, irrespective of the value
of $\lambda$.

\subsubsection*{Discussion}

In this manuscript, we derive in a rigorous way a metric that can
be used to evaluate and compare experimental high-throughput directed
evolution protocols. Like previous approaches it requires a mock experiment,
where a know mixture of functional and dysfunctional variants is submitted
to one selection round, and the change in frequency is evaluated.
However, contrary to its predecessors, our metric separates experimental
design in two categories, depending on whether they are controllable
design parameters or not. We incorporate the value of initial fraction
$p$ and the average occupancy $\lambda$ in the calculation of the
score, to provide a metric that is independent on these factors and
only reflects other, less controlled causes that may affect the efficiency
of a protocol. 

When applied to reported data, we observe that the SQI indeed provide
an informed evaluation of protocol quality. The fact that it corrects
for $p$ and $\lambda$ allows us to extract trends by comparing different
experimental approach and make hypothesis about the cause for the
observed differences in efficiency. It also clearly reveals the effect
of decreasing signal-to-background values, when the initial fraction
of active mutants becomes very small. This approach allowed to detect
abnormal selection efficiency, and link this behavior to a deviation
from the model of additive fitness, most probably due to the toxicity
of bacterial lysate. Counterintuitively, these toxic effects have
a positive effect of the selection efficiency.

The equations presented above are valid when individual mutants are
\emph{randomly} distributed in the available compartments. In this
case, one indeed obtains a poissonian distribution of the number of
mutant per compartment, on which we have based our analysis. However,
other distributions can be more relevant in some cases. For example,
some microfluidic devices use physical effects to encapsulate objects
in compartments with distribution sharper than the poisson distribution\cite{kemna_high-yield_2012,collins_poisson_2015,di_carlo_continuous_2007,edd_controlled_2008}.
Alternatively, mutants candidates can have a tendency to stick or
aggregate to each other (for instance in pairs, triples, tetrads,
etc.), which will also distort the distribution away from the poisson
law. Polyploidy represent a related case, although closer in context
to population dynamics than directed evolution protocols. Fortunately,
our mathematical approach is general enough to handle many of these
cases, and yield theoretical values of $\Delta$ in closed analytical
form, when available. 

\begin{figure}
\begin{centering}
\includegraphics[width=1\textwidth]{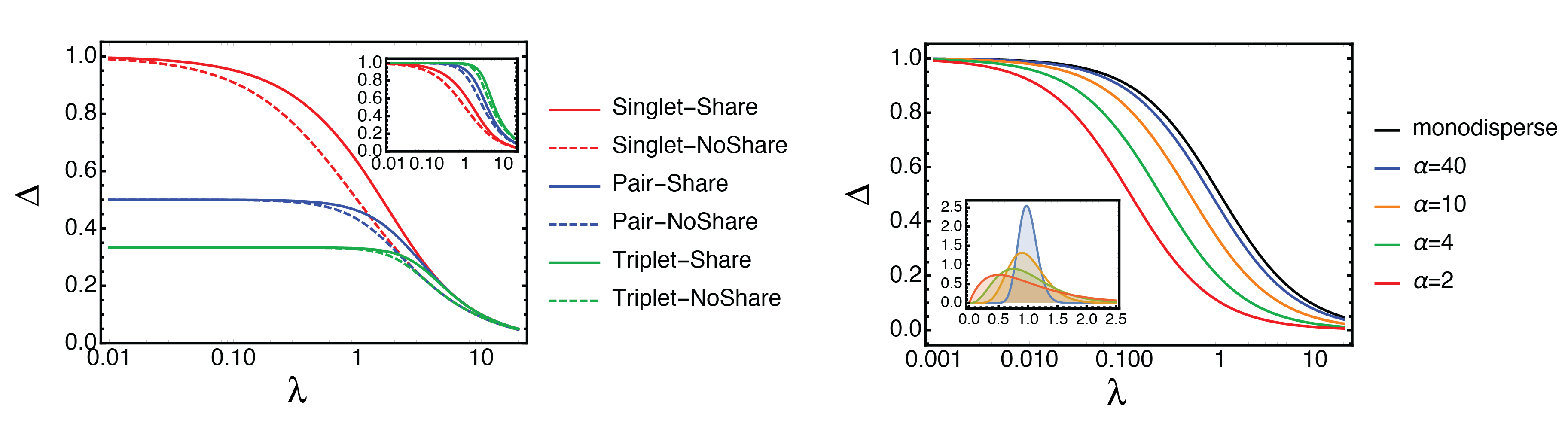}
\par\end{centering}
\caption{\label{fig:poisson deviations}Right, the effect of aggregation of
the mutants in pairs or triplets on the theoretical value of $\Delta$
(the achievable frequency jump in one round, valid for low $p$).
Full lines correspond to sharing while dotted lines indicate no sharing.
Red; encapsulation of discrete individual; Blue, encapsulation in
pairs; green, encapsulation in triplets. Inset, the rescaled curves
$\Delta/\Delta_{\lambda=0}$. Left, the effect of polydispersity of
the compartment in the case of screenings. Here we assume that the
volumes are Gamma-distributed with mean one, and various shape parameter
$\alpha$. The corresponding distributions are shown in inset.}
 
\end{figure}

The corresponding mathematical derivations are provided in SI. For
example, Fig \ref{fig:poisson deviations} shows the effect of aggregated
partitioning in pairs or triplets. When $\lambda$ is small, we recover
the intuitive result that the best possible outcome in terms of frequency
jump is the inverse of the aggregation cluster size (i.e., in the
case of $k$-clusters, the good variant can not invade more than $1/k$
of the population in one round). However, the negative effect of aggregation
gets offset when $\lambda$ increases, as all curves gather on the
same asymptote, which depends only on the sharing behavior. Interestingly,
clustered selection tends to be more resilient to reasonable increases
in $\lambda$ than unclustered selection (Fig \ref{fig:poisson deviations},
left inset). 

Many high throughput \emph{in vitro} DE protocols use emulsions as
a convenient way to provide compartments. Our bibliographic analysis
has also highlighted the role of the mono/poly-dispersity of these
emulsions on the selection process. At equal average compartment volume,
polydisperse emulsion will have more compartments containing multiple
mutants. This is expected to decrease their efficiency at selecting
one particular phenotype from a mixture. Indeed, in the present case
of selection from a active/inactive mixture of variants, we can show
that the polydisperse emulsion always perform worse than the monodisperse
setting. This effect is shown for Gamma-distributed compartment volumes,
in the case of screening, in Fig. \ref{fig:poisson deviations} right. 

In conclusion, our analysis suggests that experimentalists using high
throughput methods to search for catalysts should put their efforts
on improving the efficiency of their protocols at very low initial
fractions. Indeed, many protocols target a throughput somewhere between
$T=10^{6}$ and $T=10^{9}$ and it seems unlikely that a very rare
active mutant can be purified in a single round. However, at low $p$,
according to the equation \ref{eq:sqi}, the SQI can be written as 

\[
SQI=p'/p'_{theo}(p)\sim p'/\Delta.
\]

This means that knowing $\lambda$, and hence $\Delta$, the SQI directly
provides an estimation of the final fraction of active mutant (see
\ref{fig:p'=00003Df(p)}). Many (mostly microfluidic) approaches already
provide a good SQI at $p>10^{-3}$, implying immediate fixation in
these cases. It is therefore the ability to bring rare variants (with
an initial fraction that can be estimated as the inverse of the throughput
of the protocols) to $p'=10^{-3}$ that is going to be critical for
real applications. For example, if an SQI $>10^{-3}$ is obtained
for $p$ as low as $1/T$, then one needs only two rounds to harness
the full potential of the experimental protocol, that is, fix a functional
variant initially present as a single copy in the library.

\subsubsection*{Acknowledgements}

This work was supported by the ERC Grant ``ProFF'' (number 647275).

\bibliographystyle{plain}
\bibliography{librarypaper2}



\section*{Supplementary material}

# SI Material & Methods

**Quantifying the performance of high-throughput directed evolution protocols**

Adèle Dramé-Maigné*, Anton Zadorin*, Iaroslava Golovkova, Yannick Rondelez

## Mock selection of the KlenTaq polymerase

**Compartmentalized selection with various $\lambda$**

For mock selection experiments, an inactive version of the Klentaq polymerase was constructed via site-directed mutagenesis (Q5® Site-Directed Mutagenesis Kit from NEB) using the following primers:
    Pr_NegMut_Fwd: 5'-GGTTGCACTGGGTTATAGCCAG
    Pr_NegMut_Rw: 5'-AGCAGCCAACCTTCTTCGG
    The aspartic acid 332 (GAT) was changed into a glycine (GGT) in the DYSQIELR motif (reference motif).
    Bacteria expressing either a wild-type or the inactivated version of the KlenTaq polymerase was grown and induced at 37°C and 250 rpm for 2 hours with 0.1% of L-rhamnose. After induction, the bacteria were pelleted in a centrifuge at 5000 g for 5min. They were washed (resuspended and centrifuged 5min at 5000 g) twice in a resuspension buffer (Tris-HCl ph 7.5 50mM, NaCl 100 mM). A solution was then prepared containing Thermopol DF buffer (NEB) (1x) with 1.5mM MgSO4, 400 µg/mL of BSA9000S (NEB), 200 µM of each dNTPs (NEB), 1ng/µL of Yeast RNA (Merck), 0.4% of Pluronic F-127 (Sigma-Aldrich) and the bacteria. We add 200nM of primers:
    reverse: 5'-TTAGGTCTCACTAAGCAAAAAACCCCTC
    forward: 5'-TTAGGTCTCATCTATAATACGACTCACTATAGGGAG
    Bacteria concentration was determined using OD600 measurement. The ratio of active over inactive bacteria was 1:10. The premix was then injected in a a microfluidic device under pressure control to generate droplets of ~24µm of diameters, with bacteria at indicated $\lambda$, in fluorinated oil (Novec 7500 (Sigma-Aldrich), 2% EA (Raindance)). The size of the droplets is controlled by a flow-focusing step junction at the nozzle. The collected droplets were then transferred in a PCR tube, and the PCR was initiated: 95°C 3min, (98°C 10s, 53°C 30s, 72°C 1 min 30) x 35, 72°C 2 min.

| λ | Fraction (active/total) | | | | Active genes (fraction) enrichment (n-fold) | SQI | Bulk final | |
|---|---|---|---|---|---|---|---|---|
| | initial | error | final | error | | | fraction | error |
| **1st selection experiment** | | | | | | | | |
| 0,01 | 5,0% | | 96,0% | | 19,2 | 0.96 | 31,7% | |
| 0,1 | 7,3% | | 84,0% | | 11,5 | 0.86 | 20,7% | |
| 1 | ND* | | 80,0% | | 13,3 | 1.23 | 57,9% | |
| 5 | ND* | | 55,0% | | 9,2 | 2.52 | 24,1% | |
| **2nd selection experiment** | | | | | | | | |
| 0,01 | 20,7% | ±6,9% | 99,9% | ±0,1% | 4,8 | 1.00 | 24,63% | ±10% |
| 0,1 | 19,3% | ±4% | 97,7% | ±0,7% | 5,1 | 1.02 | 29,7% | ±4,5% |
| 1 | 16,8% | ±3,5% | 83,3% | ±2,8% | 5,0 | 1.29 | 36,6% | ±11,4% |
| 3 | 24,0% | ±6,5% | 65,9% | ±9,5% | 2,7 | 1.96 | 37,3% | ±7% |

\* Values could not be determined experimentally.
We use the experimental value of 10% for SQI calculations.

Table 1: Initial and final active fraction measured by rhqPCR

**Selectivity assay**

To assess the selectivity of the assay, rh qPCR reactions were run using IDT rhprimers and the RNaseH2. These primers allow to detect a single nucloetide mutation change. The rhPCR was set up using DreamTaq buffer (1x), 1.5mM of extra MgSO4 (NEB), 200µM of dNTPs (NEB), 1% Evagreen (Biotium), 0.5% of DreamTaq DNA polymerase (Thermofisher) and a specific amount of RNase H2 (IDT DNA technologies), 0.3 mU for the wild-type detecting reaction and 0.2 mU for the inactive Klentaq detection in 10µL reactions, and 200nM of the two adapted rhPrimers:
 forward wild-type: 5'-TTGGCTGCTGGTTGCACTGGrATTATC_3SpC3
 forward inactive: 5'-TTGGCTGCTGGTTGCACTGGrGTTATC_3SpC3
 reverse: 5'-GCTTCACGCGGAACACCAAACrATCCAC_3SpC3

The 10 µL reactions were monitored in CFX96 Touch Real-Time PCR Machine (Bio-rad) following a 3 min denaturation at 95°C and 40 standard qPCR cycles (95°C 10 s, 60°C 40s). Standards were run with plasmid of either active or inactive version of the gene starting from ~1 ng in 10 µL and diluted sequentially by 5 for 5 dilutions. To determine the proper amount of RNase H2 giving the best assay dynamic range, ranges of RNase H2 were realized for each primer couple. The results were compared to control reactions performed with the corresponding classical PCR primers.

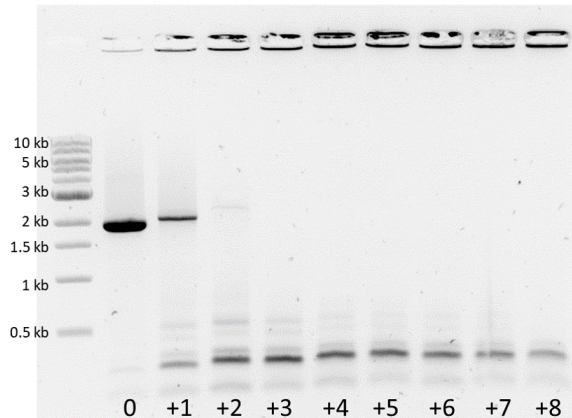

Equivalent added bacteria carrying the inactive mutant

Figure 1: Assaying the effect of droplet contamination by inactive mutants. Self-replication was run in test tubes with bacteria expressing the wild-type KlenTaq at a concentration equivalent to one bacteria per droplet (~130 bac/nL) and various amount of added bacteria expressing the mutant inactive KlenTaq (corresponding to the indicated equivalent contaminant inactive bacteria per droplet).

## Estimation of the effect of the co-encapsulation of active with inactive variants

A PCR was run in Thermopol DF buffer (NEB) (1x) with 1.5mM MgSO4, 400 µg/mL of BSA9000S (NEB), 200 µM of each dNTPs (NEB), 130 bact/nL (~1 bacteria per droplet) of bacteria expressing the active wild-type KlenTaq gene, and 200nM of reverse and forward primers:

    reverse: 5'-TTAGGTCTCACTAAGCAAAAAACCCCTC
    forward: 5'-TTAGGTCTCATCTATAATACGACTCACTATAGGGAG.

Various amount of bacteria expressing the inactive KlenTaq mutant were added to simulate the encapsulation of the active wild-type with 1 and up to 8 inactive bacteria. PCR was run using the following protocol: 95°C 3min, (98°C 10s, 53°C 30s, 72°C 1 min 30) x 35, 72°C 2 min. Results presented on 1 show that the contamination of droplets containing one active bacteria with one inactive bacteria is greatly decreasing the reaction yield. A slight amplification is visible with 2 inactive bacteria and then no gene product is observed.

# Literature review

25 articles performing high-throughput selection or screening experiment were studied and the found characteristic parameters and results are recorded in table S2 S3 and S4 according to the emulsion generation technique. All values corresponding to the initial and final quantity of the active variant were converted in ratio of active over inactive clones, when given in fraction originally.

| # | Experiment (Target) | λ | Ratio active/inactive Initial | final | Active genes enrichment (n-fold) | SQI | Achievement in xx rounds Throughput () |
|---|---|---|---|---|---|---|---|
| Screening experiments with emulsion generated by shaking or filter ||||||||
| **Several λ process : gene on beads, beads in IVTT, beads in selection droplets** ||||||||
| 1 | Directed evolution of an extremely fast phosphotriesterase by in vitro compartmentalization, 2003 (Phosphotriesterase) | 0.3 | 0.1 | 1.4 | 14 | 0.71 | 63 times higher $k_{cat}$ in 6 rounds $(6 \times 10^8)$ |
| | | | 0.01 | 0.48 | 47 | 0.41 | |
| | | | 0.001 | 0.21 | 217 | 0.23 | |
| 2 | Ribozyme-Catalyzed Transcription of an Active Ribozyme, 2011 (Ribozyme polymerase) | ? | 0.1 | ~1000** | $10^4$ | ? | Polymerase ribozyme able to synthesize a wider spectrum of RNA sequence in 13 rounds $(~5 \times 10^7)$ |
| | | | 0.001 | ~100** | $10^5$ | | |
| | | | $10^{-5}$ | ~1** | $10^5$ | | |
| **Encapsulation of cells (bacteria or yeasts)** ||||||||
| 3 | High-Throughput Screening of Enzyme Libraries: Thiolactonases Evolved by Fluorescence-Activated Sorting of Single Cells in Emulsion Compartments, 2005 (Thiolactanase) | 0.03 | ¤gating 0,0001 | 0.001 | 0.29 | 290 | 0.23 | 100-fold improvement in activity of PON1 (paraoxonase) in 6 rounds $(~5 \times 10^8)$ |
| | | | | 0.01 | 0.63 | 63 | 0.28 | |
| | | | 0,001 | 0.001 | 0.04 | 40 | 0.04 | |
| | | | | 0.01 | 0.69 | 69 | 0.42 | |
| | | | 0,01 | 0.001 | 0 | - | ? | |
| | | | | 0.01 | 0 | - | ? | |
| 4 | Ultrahigh Throughput Screening System for Directed Glucose Oxidase Evolution in Yeast Cells, 2011 (Glucose oxidase) | 0.01* | 0.1 | | 4,7 †not defined | ? | 1,8 increase in $k_{cat}$ in 1 round $(~10^7)$ |
| | | | 0.01 | | 13 | | |
| | | | 0.001 | | 33 | | |
| **Encapsulation of genes by extrusion through filter** ||||||||
| 5 | High-Throughput Screening of Enzyme Libraries: In Vitro Evolution of a β-Galactosidase by Fluorescence-Activated Sorting of Double Emulsions, 2005 (β-Galactosidase) | ? (300) | 0.33 | 1 | 3 | ? | Turn a protein of unknown function into β-Galactosidase in 2 rounds $(4 \times 10^7)$ |
| | | | 0,008 | 0.82 | 103 | | |
| | | | 0,008 enrich§ | 0.19 | 24 | | |
| | | | 0,001 | 0.16 | 160 | | |

\* not explicitly given by the authors, estimated from available data in the paper
\*\* estimated from gels given in the paper or sup mat by eyes
§ enrich is an alternative mode of screening, the other most use mode in this paper is called the purify mode
† not defined : the formula of the enrichment is not given
¤ gating : Upper % percentile in UV fluorescence gate

Table 2: Screening experiments with emulsion generated by shaking or filter

| # | Experiment | λ | Ratio active/inactive initial | Ratio active/inactive final | Active genes enrichment (n-fold) | SQI | Achievement in xx rounds Throughput () |
|---|---|---|---|---|---|---|---|
| | **Screening experiments with emulsion generated by microfluidic** | | | | | | |
| 12 | High-throughput screening for industrial enzyme production hosts by droplet microfluidics, 2014 (Amylase) | ? | 0.25 | 3.5 | 14 | ? | 2-fold increase in α-amylase production in 1 round ($10^5$) |
| 13 | A high-throughput cellulase screening system based on droplet microfluidics, 2014 (Cellulase) | ? | 0.001<br>0.014<br>0.078<br>1 | 0.43<br>1.75<br>7.62<br>21.7 | 430<br>125<br>97.7<br>21.7 | ? | Method development |
| 14 | Dissecting enzyme function with microfluidic-based deep mutational scanning, 2015 (β-Glucosidase) | 0.1 | 0.54 | 49 | 90.7 | 1.07 | Deep mutational scanning ($10^7$) |
| 15 | Ultrahigh-throughput discovery of promiscuous enzymes by picodroplet functional metagenomics, 2015 (Detection Phosphotriesterase) | 0.8 | $4.6 \times 10^{-9}$ | $2.3 \times 10^{-5}$ | $\sim 5 \times 10^3$ | $4.1 \times 10^{-4}$ | Identification of starting point sequences for new functions in 3 rounds ($10^7$-$10^8$) |
| 16 | High-throughput screening of filamentous fungi using nanoliterrange droplet-based microfluidics, 2016 (α-amylase) | 0.24 | 0.21 | 40.7 | 196 | 1.2 | 2,3 fold activity increase in 1 round ($5 \times 10^4$) |
| 17 | Ultrahigh-throughput–directed enzyme evolution by absorbance-activated droplet sorting (AADS), 2016 (Oxydoreductase) | 1 | 0.0002 | 1.27 | 6350 | 1.12 | activity >4.5-fold, kcat >2.7-fold, soluble expression 60% higher, Tm 12 °C higher in 2 rounds ($10^6$) |
| 18 | A general strategy for expanding polymerase function by droplet microfluidics, 2016 (Polymerase) | 0.1 | 0.01<br>0.001<br>0.0001 | 70.4**<br>2**<br>0.44** | ~4 500 | 1.08<br>0.8<br>0.4 | TNA polymerase in 1 round ($3.6 \times 10^7$) |

**estimated from gels given in the paper or sup mat by gel analysis with ImageJ
† not defined : the formula of the enrichment is not given

| # | Experiment | λ | Ratio active/inactive initial | Ratio active/inactive final | Active genes enrichment (n-fold) | SQI | Achievement in xx rounds Throughput () |
|---|---|---|---|---|---|---|---|
| 12 | High-throughput screening for industrial enzyme production hosts by droplet microfluidics, 2014 (Amylase) | ? | 0.25 | 3.5 | 14 | ? | 2-fold increase in α-amylase production in 1 round ($10^5$) |
| 13 | A high-throughput cellulase screening system based on droplet microfluidics, 2014 (Cellulase) | ? | 0.001<br>0.014<br>0.078<br>1 | 0.43<br>1.75<br>7.62<br>21.7 | 430<br>125<br>97.7<br>21.7 | ? | Method development |
| 14 | Dissecting enzyme function with microfluidic-based deep mutational scanning, 2015 (β-Glucosidase) | 0.1 | 0.54 | 49 | 90.7 | 1.07 | Deep mutational scanning ($10^7$) |
| 15 | Ultrahigh-throughput discovery of promiscuous enzymes by picodroplet functional metagenomics, 2015 (Detection Phosphotriesterase) | 0.8 | $4.6 \times 10^{-9}$ | $2.3 \times 10^{-5}$ | $\sim 5 \times 10^3$ | $4.1 \times 10^{-4}$ | Identification of starting point sequences for new functions in 3 rounds ($10^7$-$10^8$) |
| 16 | High-throughput screening of filamentous fungi using nanoliterrange droplet-based microfluidics, 2016 (α-amylase) | 0.24 | 0.21 | 40.7 | 196 | 1.2 | 2,3 fold activity increase in 1 round ($5 \times 10^4$) |
| 17 | Ultrahigh-throughput–directed enzyme evolution by absorbance-activated droplet sorting (AADS), 2016 (Oxydoreductase) | 1 | 0.0002 | 1.27 | 6350 | 1.12 | activity >4.5-fold, kcat >2.7-fold, soluble expression 60% higher, Tm 12 °C higher in 2 rounds ($10^6$) |
| 18 | A general strategy for expanding polymerase function by droplet microfluidics, 2016 (Polymerase) | 0.1 | 0.01<br>0.001<br>0.0001 | 70.4**<br>2**<br>0.44** | ~4 500 | 1.08<br>0.8<br>0.4 | TNA polymerase in 1 round ($3.6 \times 10^7$) |

**estimated from gels given in the paper or sup mat by gel analysis with ImageJ
† not defined : the formula of the enrichment is not given

Table 3: Screening experiments with emulsion generated by microfluidics

| # | Experiment | λ | Ratio active/inactive | | Active genes enrichment (n-fold) | SQI | Achievement in xx rounds Throughput () |
|---|---|---|---|---|---|---|---|
| | | | initial | final | | | |
| **Selection experiments with emulsion generated by shaking** | | | | | | | |
| **Encapsulation of genes** | | | | | | | |
| 19 | Man-made cell-like compartments for molecular evolution, 1998 (Methyltransferase) | 1 | 0.001 | 1 | 1000 | 0.79 | Method development ($10^{10}$) |
| 20 | Directed Evolution of Protein Inhibitors of DNAnucleases by in Vitro Compartmentalization (IVC) and Nano-droplet Delivery, 2005 (Inhibition of self-DNA destruction) | ≤1 | 0.005 | 0.33 | 66 | ? | claim >100 fold ≤ 500 fold, changing protein target in 8 rounds (~$10^{10}$) |
| | | | 0.001 | ? | | | |
| | | | 0.0004 | 0.05 | 125 | | |
| 21 | Selection of restriction endonucleases using artificial cells, 2007 (Restriction endonuclease) | 1 | 0.001 | >0.1 | ≥100 | 0.78 | 20-fold improvement of activity in 3 rounds ($10^{10}$) |
| | | | 0.01 | 1 | 100 | | |
| 22 | An in vitro Autogene, 2012 (T7 RNA polymerase mRNA) | 0.01-1 | 0.11 | 4 | 36 | ? | Too many mutations arise in 4 rounds ($10^8$-$10^{10}$) |
| **Encapsulation of bacteria** | | | | | | | |
| 23 | Directed evolution of polymerase function by compartmentalized self-replication, 2001 (Taq Polymerase) | 1.7* | 0.01-0.05 | 4.3-3.3 | 109 | 1.63 | 11-fold higher thermostability, 130-fold resistance to inhibitors in 3 rounds ($2 \times 10^9$) |
| 24 | Directed evolution of genetic parts and circuits by compartmentalized partnered replication, 2014 (tRNA synthetase) | ≤1? | 0.1 | 2.1** | 21 | ? | T7 RNA polymerase orthogonal promoter (40-60% WT promoter)+ unnatural amino acids incorporation by tryptophanyl tRNA-synthetase: suppressor tRNA (1500-fold) in 10 to 16 rounds (>$10^6$) |
| | | | 0.01 | 0.6** | 60 | | |
| | | | 0.001 | 0.05** | 50 | | |
| | | | 0.0001 | 0.033** | 330 | | |
| | | | 0.00001 | - | | | |
| | | | 0.05 | 0.27 | 5.4 | ? | |
| **Encapsulation of complexes** | | | | | | | |
| 25 | Compartmentalization of destabilized enzyme–mRNA–ribosome complexes generated by ribosome display: a novel tool for the directed evolution of enzymes, 2013 (Reverse Transcriptase) | 0.27 | 0.02 | 1 | 50 | 0.56 | 3-fold activity increase, thermoresistance in 5 rounds ($10^{10}$) |

\* not explicitly given by the authors, estimated from available data in the paper
\*\* estimated from gels given in the paper or sup mat by gel analysis with ImageJ

Table 4: Selection experiments with emulsion generated by shaking



\end{document}